# Three-dimensional echo-shifted EPI with simultaneous blip-up and blip-down acquisitions for correcting geometric distortion

**Authors:**

Kaibao Sun[1, ¶], Zhifeng Chen[2, 3, ¶], Guangyu Dan[1, 4], Qingfei Luo[1], Lirong Yan[2, 5], Feng Liu[6], and Xiaohong Joe Zhou[1, 4, 7, *]

[1]Center for Magnetic Resonance Research, University of Illinois at Chicago, Chicago, IL, United States,

[2]USC Stevens Neuroimaging and Informatics Institute, Keck School of Medicine, University of Southern California, Los Angeles, CA, United States,

[3]Department of Data Science and AI, Faculty of IT, Monash University, Clayton, VIC, Australia,

[4]Department of Biomedical Engineering, University of Illinois at Chicago, Chicago, IL, United States,

[5]Department of Radiology, Feinberg School of Medicine, Northwestern University, Chicago, IL, United States,

[6]School of Information Technology and Electrical Engineering, The University of Queensland, Brisbane, Queensland, Australia,

[7]Departments of Radiology and Neurosurgery, University of Illinois at Chicago, Chicago, IL, United States

**\*Address correspondence to:**

Xiaohong Joe Zhou, PhD; xjzhou@uic.edu; Phone: 312-413-3979; Fax: 312-355-1637

Center for Magnetic Resonance Research, University of Illinois at Chicago,

2242 West Harrison Street, Suite 103, M/C 831 Chicago, IL 60612

**Funding source:** This work was supported in part by grants from the National Institutes of Health (Grant Numbers NIH R01EB026716 and 1S10RR028898).

[¶]**Kaibao Sun** and **Zhifeng Chen** contributed equally to this work and share the first co-authorship.
**Manuscript type:** Research article

**Word Count:** ~4930

**Keywords:** 3D EPI; Blip-up and blip-down acquisitions (BUDA); Echo shifting; fMRI; Geometric distortion correction; Hankel low-rank reconstruction

The work was presented in part at the 30th Annual Meeting of the ISMRM (London, UK) in May, 2022 (Abstract No. 4390).

# Abstract

**Purpose:** Echo-planar imaging (EPI) with blip-up/down acquisition (BUDA) can provide high-quality images with minimal distortions by using two readout trains with opposing phase-encoding gradients. Because of the need for two separate acquisitions, BUDA doubles the scan time and degrades the temporal resolution when compared to single-shot EPI, presenting a major challenge for many applications, particularly functional MRI (fMRI). This study aims at overcoming this challenge by developing an echo-shifted EPI BUDA (esEPI-BUDA) technique to acquire both blip-up and blip-down datasets in a single shot.

**Methods:** A three-dimensional (3D) esEPI-BUDA pulse sequence was designed by using an echo-shifting strategy to produce two EPI readout trains. These readout trains produced a pair of k-space datasets whose k-space trajectories were interleaved with opposite phase-encoding gradient directions. The two k-space datasets were separately reconstructed using a 3D SENSE algorithm, from which time-resolved $B_0$-field maps were derived using TOPUP in FSL and then input into a forward model of joint parallel imaging reconstruction to correct for geometric distortion. In addition, Hankel structured low-rank constraint was incorporated into the reconstruction framework to improve image quality by mitigating the phase errors between the two interleaved k-space datasets.

**Results:** The 3D esEPI-BUDA technique was demonstrated in a phantom and an fMRI study on healthy human subjects. Geometric distortions were effectively corrected in both phantom and human brain images. In the fMRI study, the visual activation volumes and their BOLD responses were comparable to those from conventional 3D echo-planar images.

**Conclusion:** The improved imaging efficiency and dynamic distortion correction capability afforded by 3D esEPI-BUDA are expected to benefit many EPI applications.

**Introduction:**

Human brain mapping using fMRI is typically performed using gradient-echo echo-planar imaging (GRE-EPI) due to its fast acquisition speed and relatively high SNR per unit time [1]. However, since the sampling bandwidth is narrow (e.g., < 2 kHz) in the blipped phase-encoding direction, severe geometric distortions can occur in the presence of magnetic field inhomogeneities and other off-resonance effects [2–5]. If the associated perturbing field gradient has the same polarity as the blip phase-encoding gradient, then the images are stretched along the phase-encoding direction. Conversely, the images are compressed if the gradient polarities are opposite [1,6].

Several distortion correction techniques have been proposed for EPI images over the past decades. One approach [7] is to correct distortions in the image domain using a $B_0$-field map, which is obtained from a double-echo gradient echo (GRE) image. The method can effectively correct signal-stretching artifacts in the images. However, an EPI acquisition with only one phase-encoding polarity does not provide sufficient information to correct for signal pile-up in regions with substantial off-resonance effects, where signals from different locations are compressed into a single pixel and hence their spatial information is lost [6]. To address this issue, image-domain registration based on two EPI data sets with opposite phase-encoding directions (e.g., TOPUP in FSL) have been developed [6,8–10], and adopted in several large-scale neuroimaging studies [11,12]. Nonetheless, inaccurate spatial alignment during co-registration can lead to failures in distortion correction, especially in the anterior temporal lobe and orbitofrontal cortex under a poor SNR condition [13,14]. This issue has been elegantly addressed using a technique known as BUDA (blip-up/down acquisition) [15–19]. BUDA is based on a forward model that links the distortion-free image to the corrupted raw k-space data. This allows distortion-corrected images to be reconstructed from the k-space data acquired with blip-up and blip-down phase-encoding gradients by solving the corresponding inverse problem with Hankel structured low-rank constraint. In previous studies [15–19], BUDA has been successfully applied to susceptibility, diffusion and relaxation-weighted imaging and demonstrated its robustness in geometric distortion correction for two-dimensional (2D) and 3D EPI.

Despite its excellent performance, BUDA requires two separate acquisitions (or shots) with blip-up and blip-down, respectively, which compromises the imaging efficiency. This can be particularly problematic for fMRI because the need for two acquisitions halves the temporal resolution for resolving BOLD signal changes. Moreover, the longer scan times can also impose a physiological burden to the subject and increase vulnerability to motion. Very recently, an echo-shifting technique [20] was employed to incorporate two EPI echo-trains with the *same* phase-encoding gradient polarity into a single sequence to increase the time efficiency for 2D EPI [21,22]. This approach, however, is subject to a serious SNR loss caused by large flip angles needed in 2D multi-slice acquisitions. For example, only ~50% SNR can be retained under the optimal condition when compared to conventional 2D EPI with a 90° flip angle. In contrast, 3D acquisition typically uses a smaller flip angle for RF excitation to reduce signal saturation, and therefore fits naturally to the echo-shifting strategy. Herein we report a 3D echo-shifted EPI technique with BUDA (esEPI-BUDA) to separate two echo-trains with *opposing* phase-encoding gradient polarities in a single TR. Each echo train corresponds to an independent k-space dataset which can be interleaved between the blip-up and blip-down acquisitions. BUDA reconstruction on the two k-space datasets produces a $B_0$-field map together with a distortion-corrected image. In this study, we have demonstrated esEPI-BUDA in phantom and human brain for 3D fMRI. Since BUDA reconstruction is applied to the volume in each TR, distortion correction can be performed at each time point in an fMRI run and a series of time-resolved $B_0$-field maps can be obtained to fully capture the dynamic change.

**Methods:**

*3D esEPI-BUDA sequence:*

Figure 1 shows an example of the 3D esEPI-BUDA pulse sequence and its corresponding k-space trajectories. 3D esEPI-BUDA uses two RF pulses to acquire the blip-up and blip-down datasets in a single TR (or shot). The two signals from the two RF pulses are time-shifted and individually selected by three dephasing/rephasing gradients (or echo-shifting gradients; shaded in blue) applied along the slab-selection direction ($G_z$). The first echo-shifting gradient with an area of $(G - A)$ acts as a spoiler to dephase the transverse magnetization shortly after the first RF pulse *α*. (Note that *A* is the net area of slab-refocusing gradient associated with the first RF pulse.) After the transverse magnetization is dephased, the second RF pulse *β* is applied to excite the stored longitudinal magnetization, followed by the second echo-shifting gradient with an area of $-G$. This gradient dephases the transverse magnetization from RF pulse *β*, while, together with the net slab-selection gradient area ($2A - A = A$) associated with RF pulse *β,* rephasing the transverse magnetization produced by RF pulse *α*. The rephased signal is acquired by the first EPI echo-train with *blip-up* phase-encoding (red). The third and final echo-shifting gradient with an area of *G* is placed after the first readout echo-train to rephase the signal produced by RF pulse *β*, while dephasing the remaining transverse magnetization from RF pulse *α*. The rephased signal is acquired by the second echo-train with *blip-down* phase-encoding (green). k-Space data from both echo-trains are under-sampled (e.g., by two-fold) to shorten the echo-train length, thus enabling short and consistent TEs (e.g., 30 ms for fMRI at 3T) for both acquisitions. To make the echo times of the two echo-trains the same, the time delay between the two RF pulses ($\alpha$ and $\beta$) is determined by the center-to-center length of the two echo-trains (i.e., the summation of the third echo-shifting gradient width plus the duration of an individual echo-train).

The signal in the first echo-train is attenuated by $cos^2(\beta/2)$ [20,23,24], and thus is proportional to $M_0 \cdot \sin\alpha \cdot cos^2(\beta/2)$, where $M_0$ is the longitudinal magnetization. The signal in the second echo-train is proportional to $M_0 \cdot \cos\alpha \cdot \sin\beta$, provided that the time delay (~20 ms) between the two RF pulses is short as compared to the T1 value of the tissue. To equalize the signals for the two echo-trains, the flip angles $\alpha$ and $\beta$ need to satisfy the following condition:

$$\sin\alpha \cdot cos^2(\beta/2) = \cos\alpha \cdot \sin\beta \qquad (1)$$

The esEPI-BUDA sequence also contains a small gradient with ½ phase-encoding blip area ($G_y$) prior to the second echo-train so that the two k-space trajectories can be interleaved (Figure 1B). This allows the two k-space datasets be effectively used in the Hankel structured low-rank image reconstruction [15–19]. In principle, the esEPI-BUDA pulse sequence described above can be used for 2D or 3D imaging. In the former case, the slab-selection gradient along the $G_z$-axis is reduced to a slice-selection gradient, while in the latter case, a conventional stepping phase gradient is applied along the $G_z$-axis, providing spatial localization for both echo trains as shown in Figure 1A.

*Imaging Experiments:*

For fMRI applications, we implemented a 3D version of esEPI-BUDA on a GE MR750 3T scanner (GE Healthcare, Waukesha, Wisconsin, USA) to avoid the excessive SNR penalty that would incur in 2D implementations. Phantom and human *in vivo* experiments were performed using a 32-channel head coil (Nova Medical, Inc., Wilmington, Massachusetts, USA). All scans on healthy human volunteers were conducted with approval from the Institutional Review Board (IRB) and written informed consent from the subjects.

In the phantom experiment, a GE DQA (Daily Quality Assurance) phantom was used to validate the pulse sequence and its associated 3D image reconstruction. Phantom images from the 3D esEPI-BUDA sequence were acquired with the following parameters: TR/TE = 100/40 ms, volume TR (TR$_{vol}$) = 3.2 s (where TR$_{vol}$ is defined as the time taken to acquire a 3D volume), flip angles: α ≈ β = 15° (to satisfy Eq. (1)), FOV = 180 × 180 × 128 mm$^3$, acquisition matrix = 72 × 72 × 32, spatial resolution = 2.5 × 2.5 × 4.0 mm$^3$, under-sampling factor along the in-slab phase-encoding direction = 2 (i.e., the length of each echo-train (ETL) = 36; total length of two echo-trains = 72), and echo spacing = 0.528 ms. The corresponding k-space data from the two echo-trains were separately reconstructed using SENSE, followed by joint image reconstruction (see *Image reconstruction*). For comparison, images over the same volume were also acquired using 3D EPI (flip angle = 15°) with blip-up and blip-down acquisitions separately. To establish a reference to assess the improvement in image distortion reduction, additional images were obtained using a conventional 3D fast SPGR sequence (TR/TE = 10.6/1.4 ms, flip angle = 10°, FOV = 180 × 180 × 128 mm$^3$, and acquisition matrix = 72 × 72 × 32).

The *in vivo* experiment aimed at demonstrating the 3D esEPI-BUDA sequence for fMRI with visual stimulation. The imaging parameters were: TR/TE = 75/30 ms, TR$_{vol}$ = 2.4 s, α ≈ β = 15°, FOV = 220 × 220 × 128 mm$^3$, acquisition matrix = 72 × 72 × 32, spatial resolution = 3.1 × 3.1 × 4.0 mm$^3$, under-sampling factor along the in-slab phase-encoding direction = 2, and echo spacing = 0.456 ms. For comparison, images over the same volume were also acquired using a conventional 3D EPI sequence (flip angle = 15°) with separate blip-up and blip-down acquisitions. Similar to the phantom experiment, a conventional 3D fast SPGR image with matched FOV was also acquired as a reference (TR/TE = 10.5/1.4 ms, flip angle = 10°, FOV = 220 × 220 × 128 mm$^3$, and acquisition matrix = 72 × 72 × 32).

For the *in vivo* experiment, visual stimulation was delivered using a commercial system (SensaVue, Invivo Corporation, Gainesville, Florida, USA) with a dark-gray and light-gray checkboard pattern flashing at 8 Hz. Our block-design paradigm contained six 48 s blocks, each with a 24 s stimulation period and a 24 s rest. The total acquisition time was 4 min and 48 sec. Six healthy human subjects (age = 33.2 ± 5.9 years) were asked to fixate on the cross-hair presented at the center of the visual field during the experiment. The fMRI stimulation delivery system was integrated with an infrared camera focusing on the subject's pupil so that the subject's motion and attentiveness to the fMRI task were monitored in real time. The camera was positioned at a 90° angle with respect to the optical pathway of the visual stimulation light. A large hot mirror (35" x 17") was positioned at a 45° angle with respect to both the incident infrared and visible light beams, allowing 98% transmission of visible light while reflecting 97% of infrared.

*Image reconstruction:*

The raw k-space data acquired with the 3D esEPI-BUDA sequence were preprocessed before image reconstruction: First, phase correction was performed to remove the source for Nyquist ghosting artifacts. The zeroth- and first-order phase differences between the odd and even encodings of each echo-train were corrected based on a reference scan by setting the amplitude of the phase-encoding gradients to zero [1]. Second, to speed up image reconstruction, a model based on the principal component analysis (PCA) was used to linearly concatenate the raw data from 32 channels into 16 channels. Coil sensitivity profile was estimated using the ESPIRiT approach [25] with the data from the FOV-matched distortion-free 3D SPGR sequence. After the preprocessing, distortion-corrected images were reconstructed from the k-space data with the pipeline described in the following paragraphs.

Figure 2 shows the steps involved in 3D esEPI-BUDA image reconstruction. Each under-sampled echo-train dataset (i.e., blip-up and blip-down) first underwent 3D SENSE reconstruction individually using the following equation:

$$\tilde{I} = \underset{I}{argmin} \|UF(SI) - d\|_2^2 \qquad (2)$$

where $U$ is the sampling mask of k-space locations, $F$ represents the Fourier transform operator, $S$ is the coil sensitivity, $d$ is the k-space dataset, $I$ is the 3D SENSE-reconstructed image, and $\tilde{I}$ is the resultant $I$ after iterations. A 3D POCS (projection onto convex sets) algorithm [26,27] was employed to solve the above equation.

After obtaining the two 3D SENSE images with blip-up and blip-down encoding individually, TOPUP in FSL [9,10] was used to estimate a $B_0$-field map $E$, which was subsequently incorporated to jointly reconstruct the data from both echo-trains, as follows:

$$\tilde{I} = argmin_I \sum_{t=1}^{2} \|U_t F_t (E_t S I_t) - d_t\|_2^2 + \lambda \|\mathcal{H}(I)\|_* \qquad (3)$$

where $t$ is the echo-train index (1 or 2), $U_t$ is the sampling mask of the $t^{th}$ echo-train, $F_t$ is the fast Fourier transform operator for the $t^{th}$ under-sampled echo-train, $E_t$ is the $B_0$-field map incorporated distortion operator of the $t^{th}$ echo-train, $I_t$ is the target distortion-corrected image, and $d_t$ is the acquired under-sampled k-space data of the $t^{th}$ echo-train. The final esEPI-BUDA image was obtained by computing the average of the reconstructed blip-up and blip-down images (i.e., $I_1$ and $I_2$). The image array, denoted as $I$, is formed by concatenating $I_1$ and $I_2$. $\mathcal{H}(I)$ represents the Hankel low-rank matrix which enforces low-rankness among different echo-trains. In Eq. (3), $||.||_*$ denotes the nuclear norm of the matrix, which is the sum of singular values, and $\lambda$ is the parameter to tune the weight of structured low-rank regularization.

In 3D esEPI-BUDA joint image reconstruction, the Hankel structured low-rank constraint mitigated the phase errors between the two echo-trains and background noise based on the hypothesis that the k-space data of MR images typically have limited spatial support and/or slowly varying phase [16,17,28–30]. Herein, the Hankel matrix was constructed by consecutively selecting $9 \times 9 \times 9$ neighborhood points in k-space from each echo-train as a Hankel-block, followed by concatenating them in the column dimension. A POCS-like approach was used to solve Eq. (3) [26,27]. In the POCS iterative framework, root-mean-square error of less than 1% between two successive iterations was chosen to indicate convergency.

All image reconstructions were implemented in MATLAB (version R2019b; the MathWorks, Natick, MA, USA) for off-line reconstruction on a Linux server (CentOS, Intel (R) Core (TM) i9-7920X CPU @ 2.90 GHz and 128 GB RAM).

*Data analysis:*

To demonstrate the performance of 3D esEPI-BUDA in terms of data acquisition efficiency and distortion correction effectiveness, images from the DQA phantom and fMRI experiments were evaluated and compared between 3D esEPI-BUDA and a conventional approach with separate blip-up and blip-down acquisitions followed by a correction method using TOPUP in FSL [9,10]. This method (called 3D EPI TOPUP thereafter) performs image-domain registration between the two separately acquired datasets. Only the corrected blip-up images were retained for 3D EPI TOPUP since blip-up EPI is commonly used for fMRI data sampling while blip-down EPI serves as a reference and is acquired prior to the fMRI scan.

In the DQA phantom experiment, the horizontal low-signal block (as indicated by the green arrow in Figure 3) served as a fiducial mark to qualitatively assess the effectiveness of geometric distortion correction produced by 3D esEPI-BUDA with joint reconstruction and the two separate blip-up/down acquisitions with 3D EPI TOPUP, respectively. In the comparison, an image acquired by the 3D SPGR sequence was used as a reference. We also performed three quantitative image analyses, including structural similarity (SSIM), normalized root-mean-square error (NRMSE), and SNR to compare the performance.

In the process of image reconstruction of fMRI data, a series of $B_0$-field maps were estimated at each time point (i.e., $TR_{vol}$) as shown in Eq. [3]. The mean value and the standard deviation of $B_0$-field across the time points were calculated to evaluate the $B_0$-field time evolution. The dynamic change of the $B_0$-field map was analyzed at the frontal and occipital lobes. The resulting real-time spatial dislocation $\Delta d$ was calculated by:

$$\Delta d = \frac{2\pi \Delta f}{BW \cdot \Delta k} \quad (4)$$

where $\Delta f$ is the amount of off-resonance in Hertz that can be derived from the $B_0$-field map, $\Delta k$ is the k-space sampling interval, and *BW* is the sampling bandwidth along the blipped phase-encoding direction, which is inversely related to the echo spacing. $\Delta d$ was used to quantify the geometric distortion at each time point in the fMRI experiment.

fMRI data were analyzed using SPM8 in MATLAB. Motion correction and spatial smoothing (FWHM = 6 mm) were applied to the magnitude images, followed by statistical analyses using a general linear model for activation detection with a *p*-value threshold (FWE corrected) of < 0.05 and a spatial cluster size of at least 30 pixels. The MarsBar toolbox was used to extract and analyze the time course. Averaged time courses with standard deviations across the subjects were compared among 3D esEPI-BUDA, the two separately acquired datasets with blip-up and blip-down prior to distortion correction, and distortion-corrected images using 3D EPI TOPUP.

**Results:**

Figure 3 displays images from the phantom experiment, including 3D EPI with separate blip-up (A) and blip-down (B) acquisitions, the corresponding 3D EPI TOPUP image (C), individually SENSE-reconstructed images from the first (D) and second (E) echo train of the 3D esEPI-BUDA sequence, a resultant esEPI-BUDA image (F), and a conventional 3D SPGR image (G) as a reference. For each display, a representative image corresponding to slice No. 16 of the DQA phantom was selected from the 3D volumetric dataset. The horizontal black block in the middle of the phantom (see the green arrow in Figure 3G) was bent upwards or downwards in (A), (B), (D), and (E). Comparison of the first-row images in Figure 3 indicates that the esEPI-BUDA sequence produced high quality images (D and E) as compared with the separate acquisition strategy (A and B). The second row of images shows that both 3D EPI TOPUP and 3D esEPI-BUDA with joint k-space reconstruction effectively corrected the image distortion with similar performance. Using the distortion-free SPGR image in Figure 3G as a reference, the esEPI-BUDA image in Figure 3F exhibited a higher degree of similarity with an SSIM of 0.91 and an NRMSE of 0.06 than the image reconstructed from two separate acquisitions (Figure 3C; SSIM = 0.87; NRMSE = 0.08). Quantitative measurement of SNR revealed an approximately 42.2% increase in the 3D esEPI-BUDA image (Figure 3F; SNR = 86.4) compared with the individual 3D SENSE-reconstructed images (Figures 3D or 3E; SNR = 60.8). However, the SNR decreased by approximately 9.3% and 6.0% compared with the conventional 3D EPI images (Figure 3A or 3B; SNR = 95.3) and the resultant EPI TOPUP image (Figure 3C; SNR = 91.9), respectively.

The performance of the distortion correction on a representative human subject (26-year-old male) is demonstrated in Figure 4. Compared with the SENSE-reconstructed EPI images using the two individual echo trains (Figures 4A and 4B), esEPI-BUDA reduced the degree of image distortion and increased the SNR (Figure 4C). This becomes more apparent by analyzing a representative section selected from the 3D volume (Figure 5). Similar to the phantom results in Figure 3, each echo train in esEPI-BUDA produced good image quality as compared to the images acquired separately (first row of images in Figure 5). Both 3D EPI TOPUP on the individually acquired images and esEPI-BUDA on the simultaneously acquired images effectively corrected the dilation and compression artifacts of the images (Figures 5C and 5F), as compared to the reference SPGR image in Figure 5G.

Figure 6A shows the dynamic $B_0$-field maps in the fMRI experiment obtained using 3D esEPI-BUDA. The time evolution of the $B_0$-field maps was captured with a temporal resolution of 2.4 s (i.e., $TR_{vol}$). Each $B_0$-field map corresponded to a specific time point in the fMRI scan (numbered by 1, 2, …, 120), and the collection of the images spanned a duration of 4 min and 48 sec. Representative $B_0$-field evolutions obtained from the frontal and occipital lobes are displayed in Figure 6B. The $B_0$-field values (mean value ± standard deviation) in the two areas were -17.1 ± 1.2 Hz and 5.9 ± 1.0 Hz, respectively. During the entire time course, the maximal $B_0$-field difference (maximum value – minimum value) was approximately 6.2 Hz and 4.7 Hz for the frontal and occipital lobes, respectively. The resulting variations of spatial dislocation were estimated from Eq. (4) to be 1.95 mm and 1.47 mm (Figure 6C), respectively. These were approximately one half of the voxel size.

Results from the fMRI experiments with visual stimulation are illustrated in Figure 7, where three contiguous activation maps are overlaid on the corresponding T1-weighted images.

fMRI activations were observed in the visual cortex, as expected. The activation maps of 3D esEPI-BUDA co-registered better with the T1-weighted structural images than the two separately acquired 3D EPI images and their resultant 3D EPI TOPUP images in which the activations were beyond the border of brain parenchyma (see more discussions in the next section). Comparable average activated volumes across subjects were detected (separate blip-up acquisition: 30.6 $cm^3$, separate blip-down acquisition: 25.3 $cm^3$, 3D EPI TOPUP: 30.7 $cm^3$, and 3D esEPI-BUDA with joint reconstruction: 27.2 $cm^3$). The average time courses across the six subjects at the visual cortex are shown in Figure 8, where the 3D esEPI-BUDA sequence and the sequences with separate acquisitions produced similar BOLD signal changes (~3%), which was higher than 3D EPI TOPUP (~2%). The decreased signal change in 3D EPI TOPUP was likely caused by the data interpolation in the TOPUP tool of FSL. In general, an excellent consistency was observed in the averaged activation volume and the BOLD signal change between 3D esEPI-BUDA and 3D EPI TOPUP.

**Discussion:**

We have demonstrated a technique – esEPI-BUDA in which two interleaved k-space datasets with reversed k-space trajectories can be acquired in a single shot. Compared to traditional BUDA techniques for image distortion correction [15–19], esEPI-BUDA offers a major advantage by improving the acquisition efficiency. With concurrent blip-up and blip-down acquisitions in a single shot, esEPI-BUDA can also be more resilient to motion and allow $B_0$-field maps to be estimated dynamically throughout a time series. The latter is particularly valuable as the $B_0$-field maps can be incorporated into the forward joint parallel imaging reconstruction model with Hankel structured low-rank constraint to dynamically correct the geometric distortions.

As an attractive technique for distortion correction in EPI, BUDA has been successfully applied to susceptibility-weighted imaging, diffusion-weighted imaging, and T2 mapping [15–19].

Extension to fMRI has been challenging as the acquisitions of blip-up and blip-down data in two separated TRs lead to a substantially degraded temporal resolution (i.e., > 4 s). By utilizing the echo-shifting strategy [20,31], esEPI-BUDA incorporates two echo trains into a single pulse sequence, hence overcoming the temporal resolution limitation as demonstrated by the short sequence TR (i.e., 75 ms) and short volume TR (i.e., 2.4 s) achieved in our fMRI experiments. A possible counter argument is that the blip-up and blip-down datasets could be acquired sequentially in two successive TRs with each TR being one half of the TR (e.g., 37.5 ms) used in 3D esEPI-BUDA. However, such an approach would lead to a considerably shorter TE, hence compromising the BOLD sensitivity. To maintain an adequate BOLD sensitivity with a TE of approximately 30 ms at 3 Tesla, the sequence TR would need to be 108 ms, which is substantially longer than the TR (i.e., 75 ms) achieved in 3D esEPI-BUDA. The time-savings afforded by 3D esEPI-BUDA arise from two primary sources. Firstly, the time elapsed between the second radiofrequency pulse and the onset of the second echo train is avoided. Secondly, redundant fat saturation RF pulse is eliminated because the two echo trains in esEPI-BUDA share a single fat saturation pulse at the beginning of each TR.

In 3D esEPI-BUDA with the echo-shifting strategy, a relatively long TE can lead to signal loss in the frontal area where the $B_0$-field inhomogeneity is substantial. This is evident in Figure 4 where a dark hole in the frontal area was observed in the seven slices of the second row and the first two slices of the third row. To confirm the cause of this phenomenon, we carried out an auxiliary experiment utilizing a conventional 3D BUDA sequence with a minimal TE of 12.8 ms and compared the results with those from 3D esEPI-BUDA with TE = 30 ms. As shown in Figure S1 of the Supplementary Materials, the appearance of the dark hole strongly depends on TE, as expected. When TE was reduced from 30 ms to 12.8 ms, the dark hole became much less visible.

Owing to the pair-wise acquisition of the blip-up and blip-down datasets, esEPI-BUDA is capable of capturing real-time $B_0$-field temporal variations in an fMRI scan. The temporal variations can be a reflection of subject motion [2], physiological respiration [32,33], heating of the gradient system [34], mechanic vibration [34], and/or other system instabilities. Although TOPUP in FSL [9,10] can effectively correct distortion at a single time-point (Figures 3C and 5C), distortion correction based on retrospective estimation of the $B_0$-field map cannot account for the dynamic off-resonance effects. In contrast, esEPI-BUDA is capable of dynamically estimating the $B_0$-field maps, even when temporal instability changes rapidly. This allows for real-time image distortion correction (Figure 6). In our fMRI experiments, we did not observe a substantial difference in $B_0$ maps between two adjacent temporal points, which could be attributed to the short TR and the excellent temporal stability of our MRI scanner. However, should rapid $B_0$-field fluctuation occur, the proposed method can be deployed to mitigate the problem.

In parallel to the substantial scan time reduction by using the echo-shifting strategy, we also achieved a short echo-train length for the individual blip-up and blip-down acquisitions by employing parallel imaging in the $G_y$-direction with a two-fold acceleration. The use of parallel imaging also contributed to reduced image distortion (Figures 5A and 5B vs. Figures 5D and 5E). Higher acceleration factors can be potentially achieved by applying data under-sampling in both $k_z$ and $k_y$ directions to reduce not only the echo-train length but also the volume TR. An example is to employ 2D CAIPIRINHA (controlled aliasing in parallel imaging results in higher acceleration) [35,36] to accomplish multi-dimensional acceleration. The feasibility of using higher acceleration factors and temporal resolutions in esEPI-BUDA will be further explored in future studies.

In the esEPI-BUDA sequence, the blip-up and blip-down echo-trains were interleaved so that complementary k-space datasets could be used in the joint reconstruction. The joint reconstruction can significantly reduce the g-factor noise penalty when compared with reconstructions on the blip-up and blip-down acquisitions separately [37,38]. In this study, Hankel structured low-rank regularization was also exploited to strengthen the joint of the blip-up and blip-down datasets by a low-rank constraint, which not only reduces the g-factor, but also removes the phase errors between different echo-trains without the need of navigation [16]. It is worth noting, however, that the echo-shifting strategy results in a theoretical SNR loss of $(1 - cos^2(\beta/2))$ [20]. For our phantom studies, the theoretical SNR loss was only ~1.7% because of a low flip angle ($\beta$ =15°). The experimentally measured SNR loss (9.3%), however, was higher, likely caused by the noise amplification during the image reconstruction.

In Figure 7, it was observed that some activated voxels occurred outside the brain only for the conventional sequences, but not for esEPI-BUDA. This was likely due to the varying degree of undesirable contribution of draining veins. In esEPI-BUDA, the signals from the vein were most likely suppressed because of the accentuated intra-vessel magnetization dephasing imposed by the additional gradient pairs. Consequently, the esEPI-BUDA sequence can improve spatial correspondence between functional activations and the underlying brain parenchyma in comparison to the conventional approach.

The 3D esEPI-BUDA technique and our study have limitations. First, the echo-shifting strategy limits the shortest TE achievable in esEPI-BUDA. Assuming that partial Fourier k-space encoding is not employed, the shortest TE is given by $0.5 \times T_{ss} + 2 \times T_{es} + 1.5 \times esp \times ETL$, where $T_{ss}$ is the duration of slab-selection gradient, $T_{es}$ is the duration of the echo-shifting gradient, $esp$ is the echo spacing in the echo-train, and $ETL$ is the echo-train length of each echo-train. Our analysis showed that the minimum TE of ~28 ms could be achieved in the fMRI experiment, which was fortunately adequate for BOLD contrast. A lengthy echo-train needed by high spatial resolution, however, can increase the minimum TE in the 3D esEPI-BUDA sequence. An excessively long TE reduces the SNR and increases the sensitivity to magnetic field inhomogeneities and flow effects. This issue can be mitigated by reducing the duration of the echo-shifting gradients and/or incorporating parallel imaging with a higher acceleration factor to shorten the echo-trains. The latter, however, can decrease the SNR [39–41], exacerbate residual aliasing artifacts [39–41], and compromise BOLD detectability [42,43]. Second, esEPI-BUDA employed two RF pulses within a single TR. The short interval between the two RF pulses could cause signal saturation in tissues with relatively long T1, such as the gray matter. This can alter the tissue contrast in esEPI-BUDA images, as evidenced by the reduced contrast between the white and gray matters (Figure 5F). However, such saturation effect did not compromise the BOLD contrast. Therefore esEPI-BUDA remains robust for fMRI studies (Figure 7). Third, although the concept of esEPI-BUDA can be applied to both 2D and 3D acquisitions, our study has focused on the 3D implementation because 2D esEPI-BUDA would require a large flip angle of the RF excitation pulses, which compromises the overall SNR in echo-shifted acquisitions. Such an SNR loss may be compensated for at a higher magnetic field (e.g., 7 Tesla). Lastly, our neuroimaging demonstration was limited to fMRI with a simple visual paradigm. The esEPI-BUDA technique can also be extended to other EPI-based neuroimaging applications such as diffusion imaging [44] and perfusion imaging, which should be explored in the future.

**Conclusion:**

We have demonstrated a novel technique, esEPI-BUDA, to enable distortion-corrected whole-brain 3D fMRI without increasing the scan time. The proposed method integrates the blip-up and blip-down data acquisitions in a single shot, followed by joint reconstruction. The integrated data acquisition strategy also produces time-resolved $B_0$-field maps that can be incorporated into image reconstruction to achieve dynamic image distortion correction. The esEPI-BUDA technique has been demonstrated on the phantom and human brain. Distortion-corrected 3D echo-planar images are successfully obtained with adequate SNR and BOLD sensitivity in a task-based fMRI study. With these demonstrations, esEPI-BUDA is likely to find other neuroimaging applications such as 3D/2D fMRI, diffusion imaging, and perfusion imaging.

**Acknowledgements:**

This work was supported in part by the National Institutes of Health (NIH) under award numbers 5R01EB026716 and 1S10RR028898. ZC and LY are supported by NIH Grants R01NS118019 and RF1AG072490. The content is solely the responsibility of the authors and does not necessarily represent the official views of the NIH or the other funding agencies. The authors are grateful to Drs. Kezhou Wang, Alessandro Scotti, and Muge Karaman for helpful discussions.

**Data availability statement**

The sample data and reconstruction code used in this study will be made available at https://github.com/uic-cmrr-3t/esEPI from the date of publication. Other relevant image datasets will be released upon request to the corresponding author, subject to local IRB and other regulations.

**Figures**

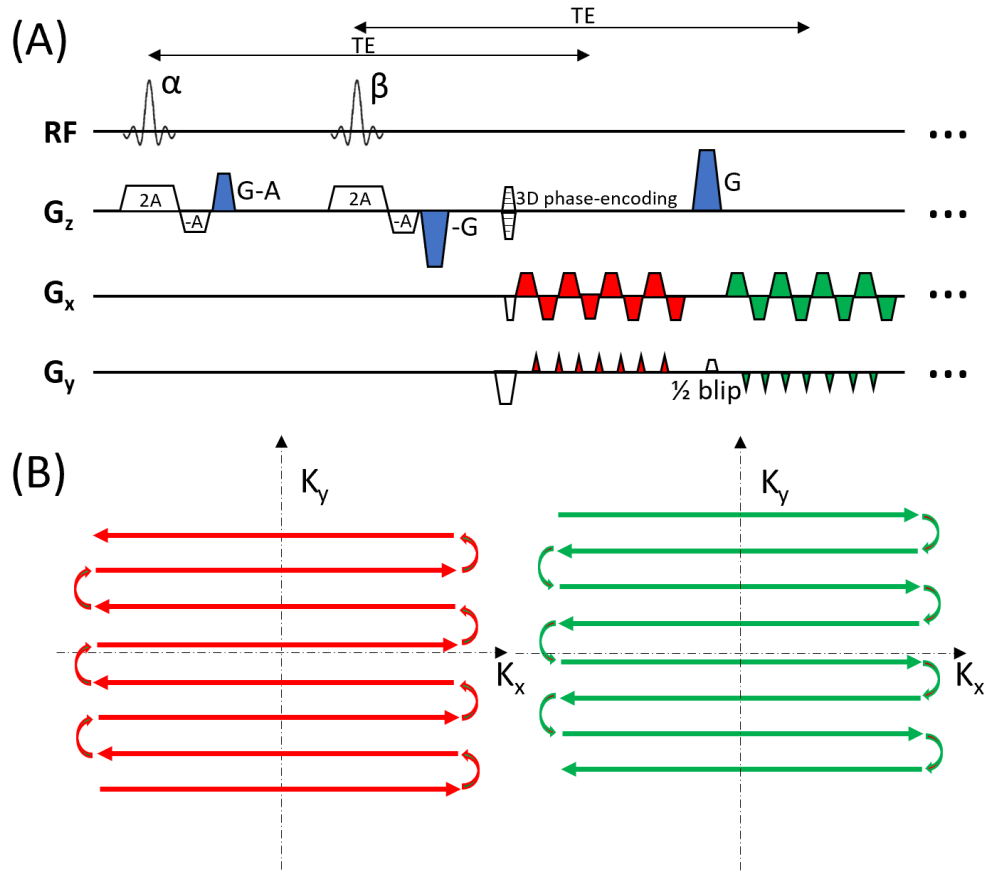

**Figure 1:** A schematic to illustrate 3D echo-shifted EPI with blip-up and blip-down acquisitions (esEPI-BUDA) in a single TR (A), and the corresponding two k-space trajectories (B). Echo-shifting gradients (shaded in blue) are applied along the slab-selection direction ($G_z$) to select the signals for the two echo-train acquisitions (red and green) resulting from the first and second RF pulses, respectively. A small gradient with ½ phase-encoding blip area ($G_y$) is played out prior to the second echo-train so that the two k-space trajectories are interleaved as shown in (B). The sequence can be readily modified for 2D multi-slice imaging by changing the slab-selection gradient to slice-selection gradient while eliminating the stepping phase-encoding gradient in $G_z$.

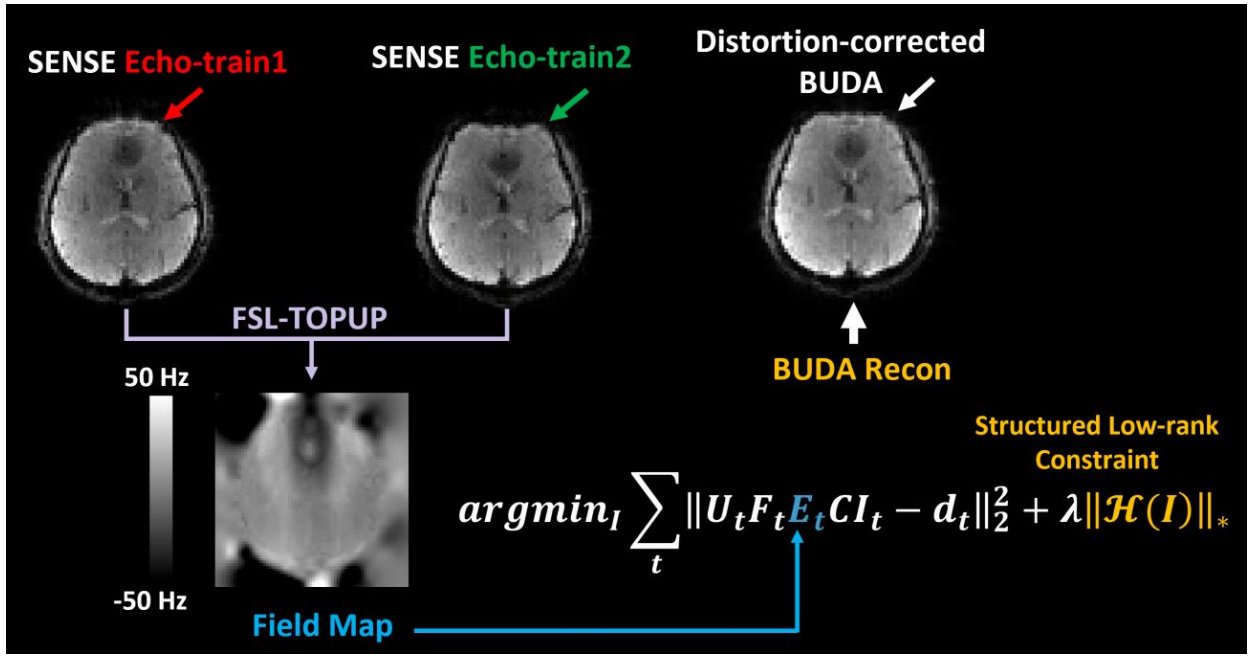

**Figure 2:** Illustration of the image reconstruction steps involved in esEPI-BUDA by using a section of the human brain image as an example. Each under-sampled echo-train dataset (i.e., Echo-train 1 with blip-up acquisition and Echo-train 2 with blip-down acquisition, as shown in Figure 1B) first underwent 3D SENSE reconstruction individually, followed by TOPUP in FSL to estimate a $B_0$-field map $E$. The $B_0$-field map was subsequently incorporated to jointly reconstruct the data from both echo-trains with Hankel structured low-rank regularization for geometric distortion correction.

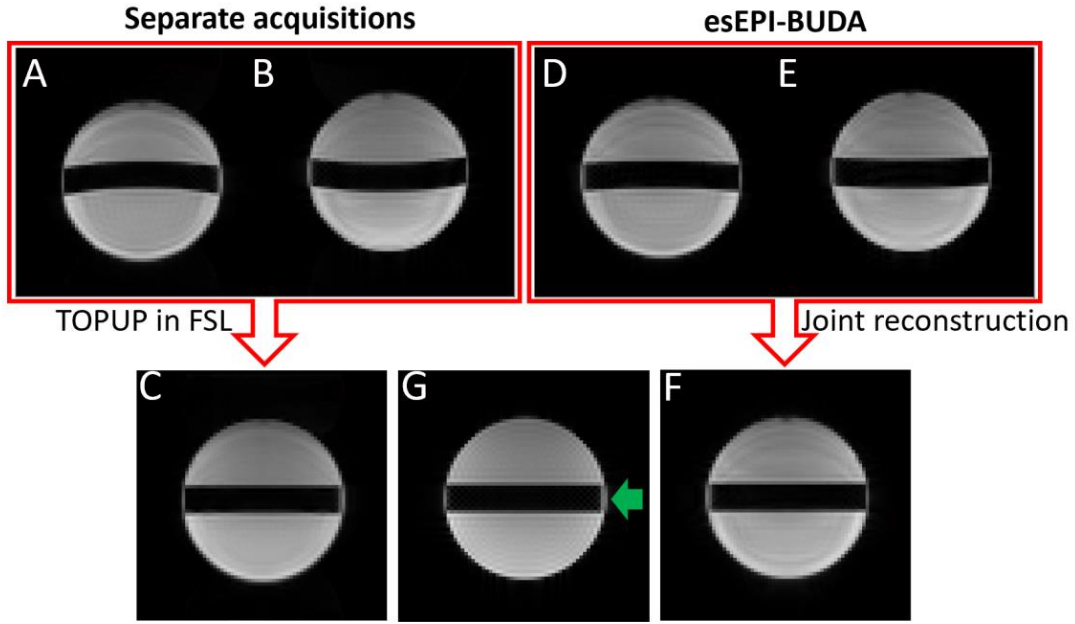

**Figure 3:** Representative images of a section selected from the 3D datasets of the DQA phantom acquired using 3D EPI with separate blip-up (A) and blip-down (B) acquisitions, the corresponding 3D EPI TOPUP image (C), individually SENSE-reconstructed images from the first (D) and second (E) echo train of the 3D esEPI-BUDA sequence, the resultant esEPI-BUDA image (F) with joint reconstruction, and a conventional 3D SPGR image (G). Image distortion was substantially corrected in (C) by using TOPUP in FSL, and in (F) by using esEPI-BUDA which jointly reconstructed an image from two interleaved k-space datasets with reversed k-space trajectories in one single shot. Using the distortion-free SPGR image as a benchmark, the esEPI-BUDA image in (F) demonstrated a greater similarity with a SSIM of 0.91 and a NRMSE of 0.06 than the EPI TOPUP image which yielded an SSIM of 0.87 and an NRMSE of 0.08.

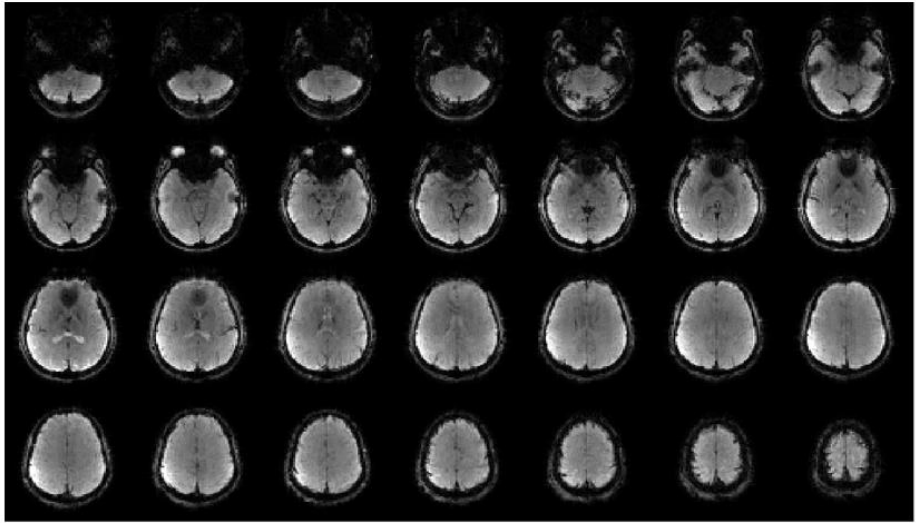

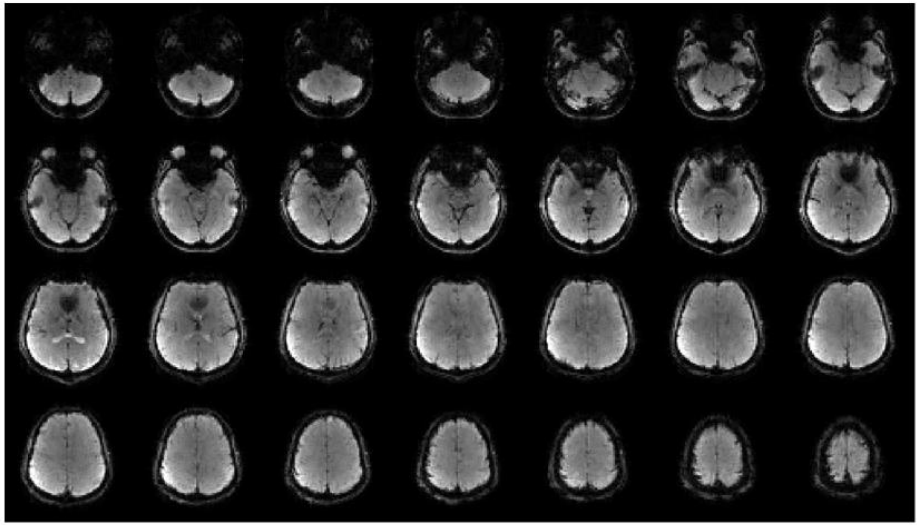

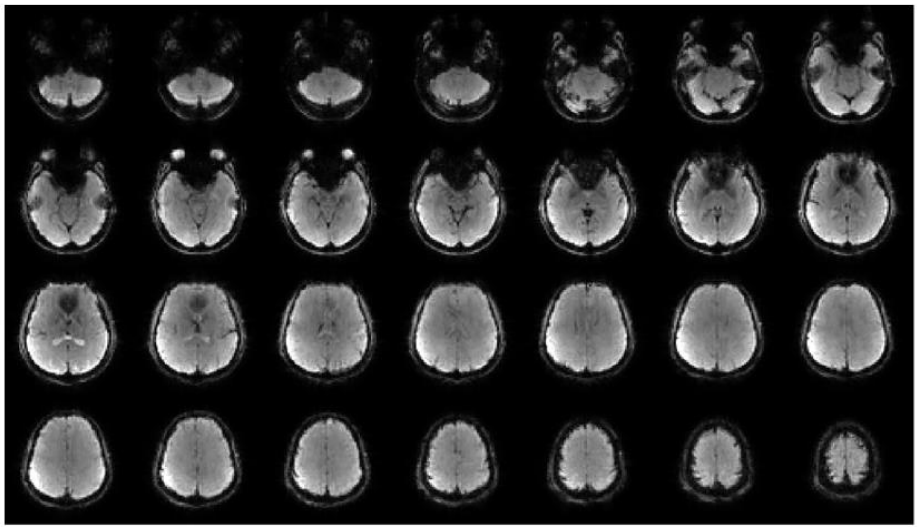

**Figure 4:** Representative whole-brain 3D images from a healthy human subject. (A) and (B): SENSE-reconstructed images using the first and the second echo train in Figure 1A, respectively; (C): the corresponding 3D esEPI-BUDA images from both echo trains acquired in a single scan.

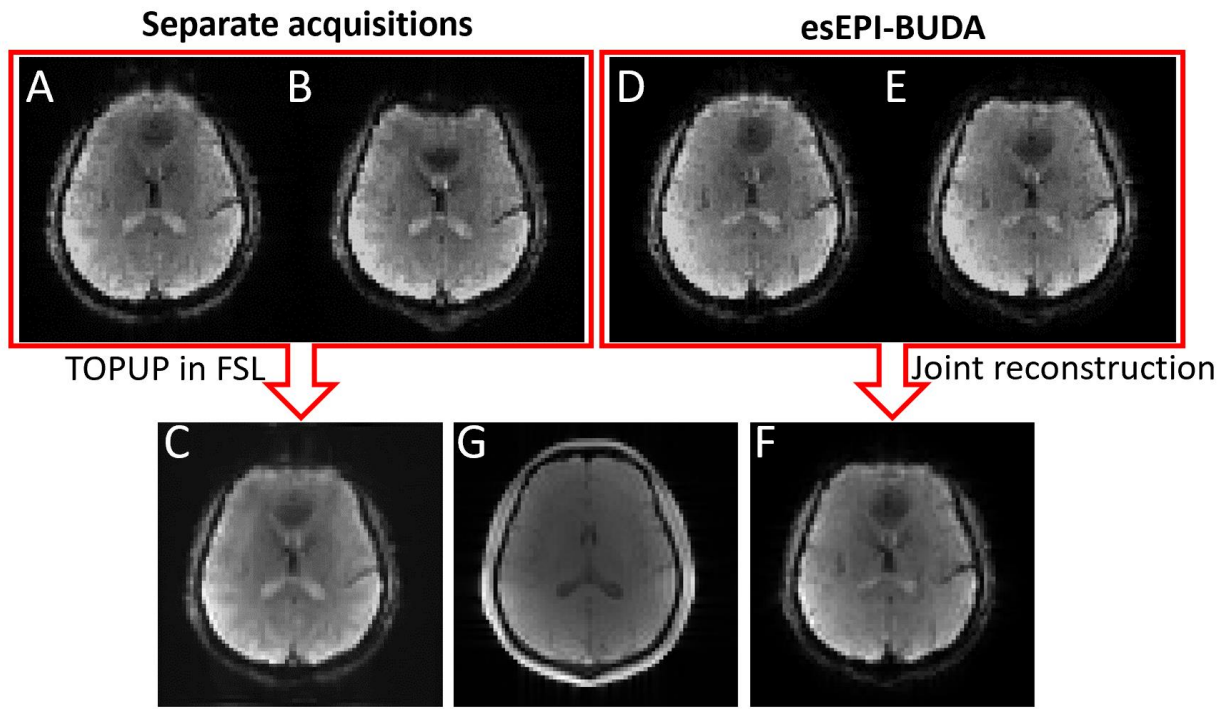

**Figure 5:** Representative images of a slice with a voxel size of 3.1 × 3.1 × 4 mm³ selected from the 3D datasets of the human brain. (A) and (B) were acquired using 3D EPI with separate blip-up (A) and blip-down (B) acquisitions. (C) displays an image reconstructed from (A) and (B) by using 3D EPI TOPUP. (D) and (E) were reconstructed from the first and second echo train of the esEPI-BUDA sequence, respectively. (F) shows the resultant esEPI-BUDA image reconstructed from both echo trains. (G) displays a conventional 3D SPGR image. Image distortion was effectively corrected in (C) and (F).

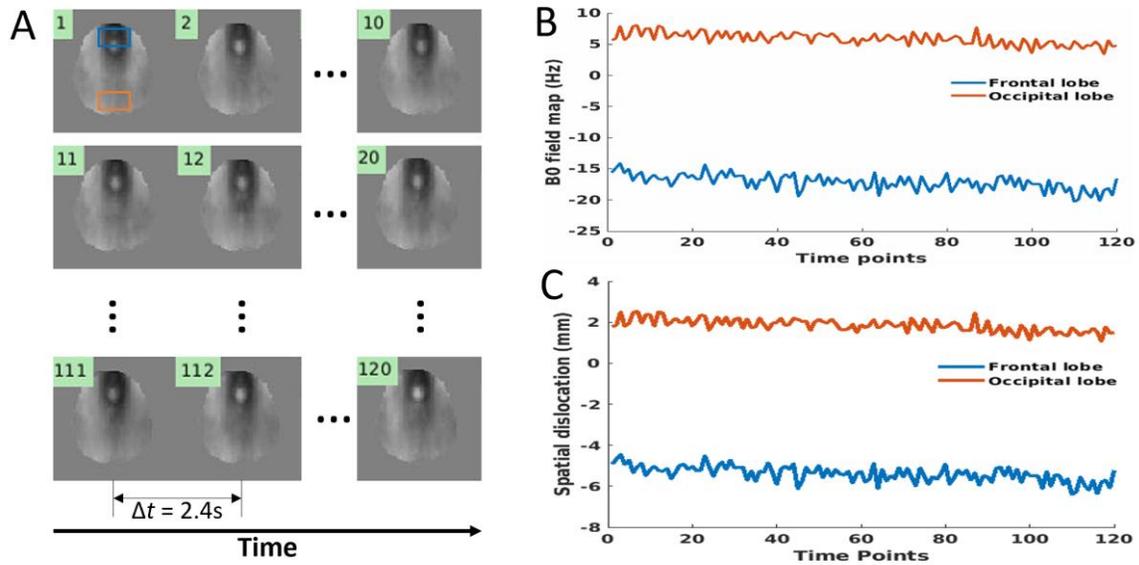

**Figure 6:** A set of 120 $B_0$-field maps (A) in a slice arbitrarily selected from the 3D datasets of the human brain, covering a total time span of 4 min and 48 sec with a temporal resolution of 2.4 sec. The dynamic $B_0$ evolutions at the frontal (blue box) and occipital (orange box) lobes are displayed in (B). The corresponding spatial dislocation caused by $B_0$ are shown in (C), as calculated from Eq. (4). The maximal $B_0$ shifts during the process were approximately 6.2 Hz and 4.7 Hz in the two brain regions, corresponding to the spatial dislocations of 1.95 mm and 1.47 mm, respectively.

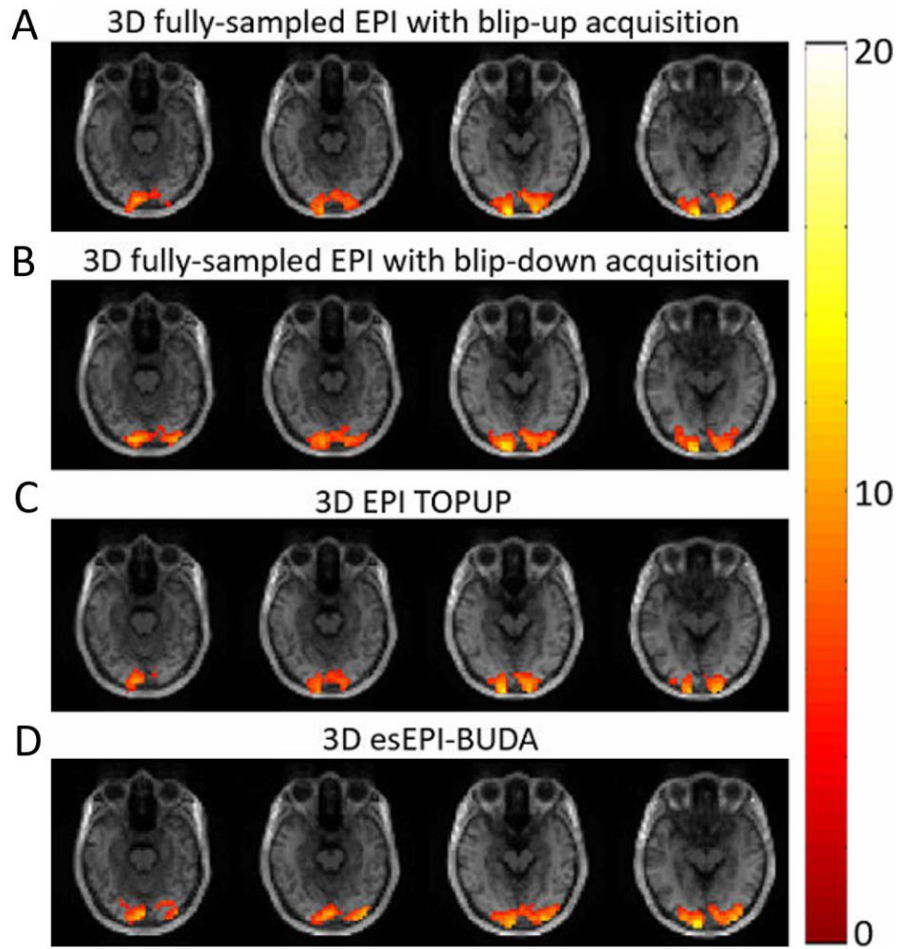

**Figure 7:** Representative fMRI visual activation maps of 3D EPI with separate blip-up (A) and blip-down (B) acquisitions, the corresponding 3D EPI TOPUP (C), and 3D esEPI-BUDA (D), overlaid onto the T1-weighted MP-RAGE images. The activation maps in (D) showed the best spatial correspondence with the brain parenchyma in the T1-weighted structural images, when compared with the activation maps in (A) – (C). The color bar indicates the scale of t-values.

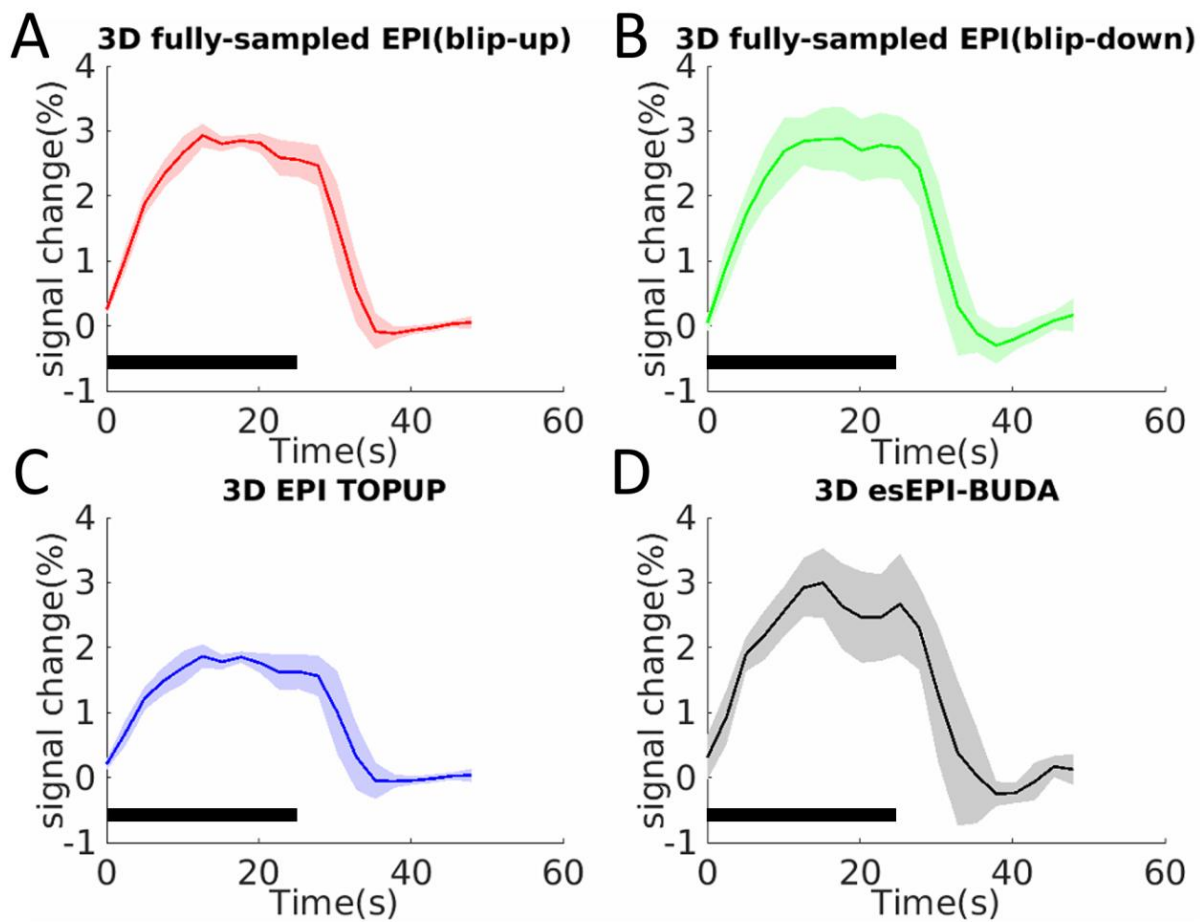

**Figure 8:** The averaged time course with standard deviations across the six subjects from the 3D EPI with separate blip-up (A) and blip-down (B) acquisitions, the corresponding 3D EPI TOPUP (C), and the 3D esEPI-BUDA sequence (D). Both 3D esEPI-BUDA and conventional 3D EPI produced similar BOLD signal change (~3%), which was higher than 3D EPI TOPUP (~2%). The black bar in each sub-figure represents the visual stimulation time period, which was 24 sec.

**Supporting Figure S1**

**Three-dimensional echo-shifted EPI with simultaneous blip-up and blip-down acquisitions for correcting geometric distortion**

In order to analyze the effects of echo time (TE) on the signal loss caused by the $B_0$-field inhomogeneity in the frontal lobe area, we performed an auxiliary experiment utilizing a conventional 3D BUDA sequence with one half of the TR (37.5 ms) and a minimal TE of 12.8 ms as a reference and compared the results with those from 3D esEPI-BUDA with TR of 75 ms and TE of 30 ms. The results are shown in Figure S1, where the appearance of the "dark hole" in some sections (see the arrows in Figure S1 A) strongly depends on TE as expected. The dark hole became much less visible when TE was reduced from 30 ms to 12.8 ms.

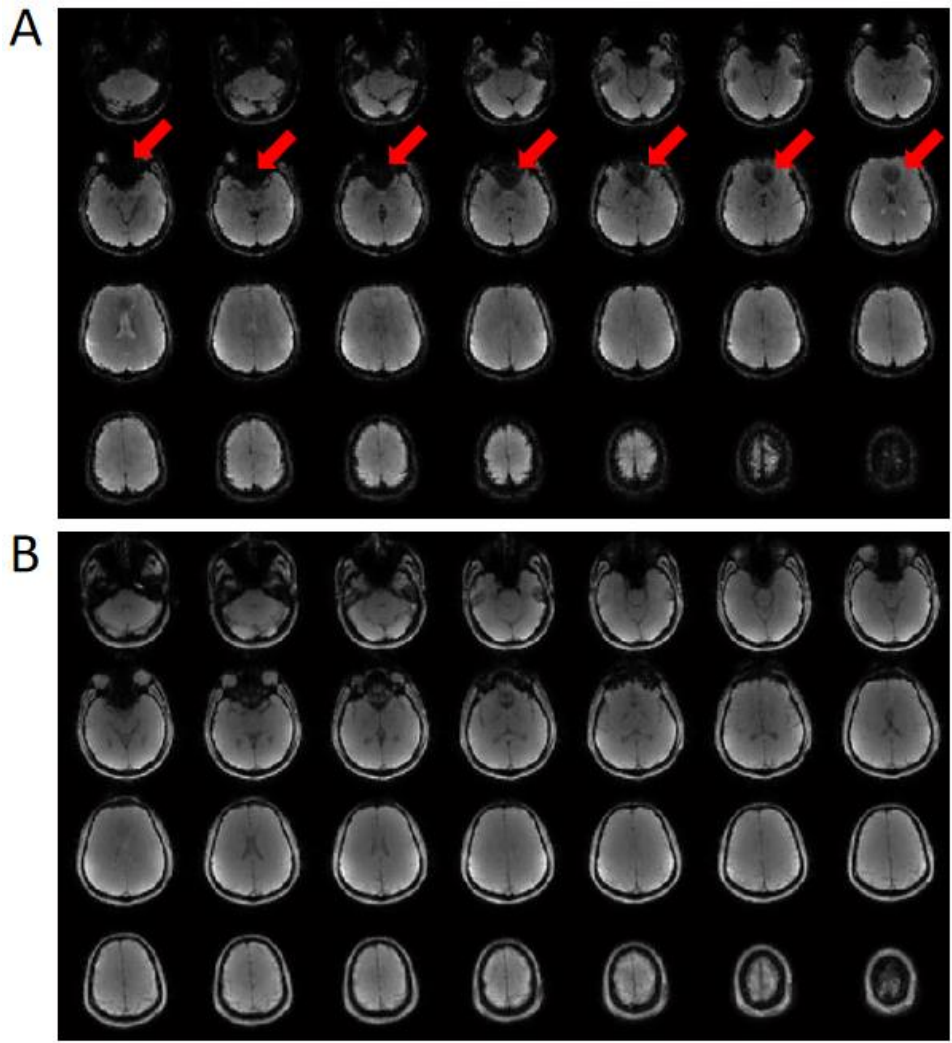

Figure S1: Signal loss caused by the $B_0$-field inhomogeneity depends on the echo time (TE). The size of the dark hole (see the arrows) at the frontal area increases with TE. (A): esEPI-BUDA images with a TE of 30 ms; (B): conventional BUDA images with a TE of 12.8 ms.